\documentclass[%
 reprint,
 superscriptaddress,
 amsmath,amssymb,
 floatfix,
 longbibliography,
 aps,
 prapplied
]{revtex4-2}

\usepackage{mathtools}
\usepackage{xcolor}
\usepackage{graphicx}
\usepackage[colorlinks,urlcolor=blue,citecolor=blue,linkcolor=blue]{hyperref}
\hypersetup{
  pdftitle={Higher-Order Kuramoto Oscillator Network for Dense Associative Memory},
  pdfauthor={Jona Nagerl and Natalia G. Berloff}
}

\begin{document}

\title{Higher-Order Kuramoto Oscillator Network for Dense Associative Memory}
\author{Jona N\"agerl}
\affiliation{Department of Applied Mathematics and Theoretical Physics, University of Cambridge, Wilberforce Road, Cambridge CB3 0WA, United Kingdom}
\author{Natalia G. Berloff}
\email[Contact author: ]{N.G.Berloff@damtp.cam.ac.uk}
\affiliation{Department of Applied Mathematics and Theoretical Physics, University of Cambridge, Wilberforce Road, Cambridge CB3 0WA, United Kingdom}
\date{\today}

\begin{abstract}
Networks of phase oscillators can serve as dense associative memories when they incorporate genuine many-body coupling beyond the classical Kuramoto model's pairwise interaction. Here we introduce a generalized Hebbian Kuramoto model that combines a conventional two-body, first-Fourier-harmonic coupling with a genuine four-body phase interaction, inspired by dense Hopfield memory theory. For identical oscillators with Langevin noise, equilibrium mean-field theory yields a phase diagram with a tricritical point when the four-body coupling is three times the pairwise coupling, where continuous retrieval onset gives way to a discontinuous, hysteretic transition. We separately analyze a deterministic model with Lorentzian frequency disorder using the Ott--Antonsen ansatz. In that model the onset changes from supercritical to subcritical when the four-body coupling is six times the pairwise coupling, and the two descriptions agree only at the linear instability threshold. In the four-body-dominated equilibrium regime, stored-pattern and incoherent states coexist. We determine the bistable region and the free-energy barriers for transitions in both directions, and show that the noise-induced memory-loss time grows exponentially with the number of oscillators, with a rate set by the relevant barrier. At finite memory load, a phenomenological cavity signal--to--noise closure models pattern crosstalk, while simulations measure finite-size strong-cue recovery and retention thresholds. Nominal power-law fits over the simulated size range have exponents above one in several four-body-dominated cases, with the largest fitted exponent in the purely four-body zero-temperature data; these fits are not asymptotic capacity laws. These results connect higher-order Kuramoto synchronization with dense Hopfield memories and identify the additional ingredients needed to implement the model in oscillator hardware.
\end{abstract}

\maketitle
\section{Introduction}\label{sec:intro}
Associative memory recovers a stored pattern from a partial or corrupted cue. Hopfield's fully connected binary-spin model is the standard example \cite{hopfield1982neural}. With pairwise couplings its storage capacity is $O(N)$, and spurious minima proliferate beyond capacity.

To overcome these limitations, dense associative memory (DAM) models introduce higher-order interactions into the energy landscape,
\begin{equation}
E(\boldsymbol{\sigma})
  =
  -\sum_{m=1}^{p}
   \frac{1}{m!\,N^{m-1}}
   \sum_{i_{1},\dots,i_{m}=1}^{N}
   J^{(m)}_{i_{1}\dots i_{m}}\,
   \sigma_{i_{1}}\cdots\sigma_{i_{m}},
\end{equation}
where $\sigma_i\in\{-1,+1\}$ are Ising spins, $p$ is the largest retained interaction order, and $J^{(m)}_{i_{1}\dots i_{m}}$ is a symmetric order-$m$ coupling tensor. With Hebbian couplings,
\begin{equation}
    J^{(m)}_{i_{1}\dots i_{m}}\!
   =\!
     \sum_{\mu=1}^{P}
     \xi_{i_{1}}^{\mu}\!\cdots\xi_{i_{m}}^{\mu},
     \label{eq:originalhebbsrule}
\end{equation}
where $\{\xi^{\mu}\}_{\mu=1}^{P}$ are the binary patterns to
be stored. The factor $N^{-(m-1)}$ is written explicitly in the energy so that every interaction order contributes an extensive $O(N)$ term; equivalently, this normalization can be absorbed into the definition of $J^{(m)}$. For a polynomial overlap energy of degree $p$, signal--to--noise estimates give a fixed-error storage scale $P_{\max}=O(N^{\,p-1})$; imposing perfect recovery introduces an additional logarithmic reduction. Thus for $p>2$ the fixed-error scale is superextensive, while the precise capacity depends on the retrieval criterion \cite{krotov2016dense,demircigil2017associative}. This class of models has smooth degradation at overload and forms a bridge to modern energy-based learning \cite{momeni2025training}; for suitable choices of the interaction potential, dense Hopfield energies can be related to the objectives of multilayer networks \cite{krotov2016dense,demircigil2017associative}.

Oscillator networks provide a continuous-phase alternative to binary memories. Relative phases encode a pattern, while retrieval corresponds to attraction toward a phase-locked configuration. Because the overlap is continuous, it also provides a graded diagnostic of proximity to a stored phase pattern.

The Kuramoto model, with its simple sinusoidal (first-harmonic) coupling, provides a canonical framework for studying synchronization \cite{acebron2005kuramoto}. In the binary-pattern constructions analyzed in Refs.~\cite{nishikawa2004oscillatory,aonishi1998phase}, pairwise first-harmonic coupling alone supports strict error-free retrieval of only a small number of patterns.
One remedy is to add a second Fourier harmonic, $\sin(2\phi_{ij})$, to the pairwise coupling. This second-resonance term creates two preferred phase differences, stabilizes binary phase patterns, and permits storage on the order of $N/\log N$ under the error-free retrieval criterion analyzed in Ref.~\cite{nishikawa2004oscillatory}.

The second-resonance model and dense associative memories suggest a different extension: genuine $p$-body phase interactions. Three- and four-body couplings can produce multistability and discontinuous synchronization transitions that are absent from purely pairwise first-harmonic dynamics \cite{skardal2020higher,gambuzza2021stability}.

We study $p=4$, the lowest even order above two in the polynomial overlap construction used here. It contains equal numbers of positive and negative phase factors and therefore preserves the global $U(1)$ symmetry \cite{demircigil2017associative}. The two channels play different roles: the pairwise term controls the linear instability of the incoherent state, whereas the four-body term changes the nonlinear branches and their barriers. Higher-order phase terms also arise in phase reductions of $S_N$-equivariant systems of weakly coupled limit-cycle oscillators \cite{ashwin2016higher}.

Nonlinear optical mixing, condensate networks, superconducting plaquette couplers, and oscillator-based optimization machines provide possible ingredients for effective many-body phase coupling \cite{kumar2020large,salem2008signal,min2009near,rose2013circular,sharping2002optical,kalinin2019polaritonic,stroev2021discrete,kanao2021high,bashar2023designing,bashar2023oscillator,nikhar2024all}. None of these platforms is claimed to realize Eq.~\eqref{eq:high-order-kuramoto} as written. Such a device would also require programmable signed Hebbian weights and the phase combination $\theta_j+\theta_k-\theta_\ell-\theta_i$.

We analyze two distinct microscopic models. The identical-oscillator Langevin model is treated by equilibrium mean-field theory, while the deterministic model with Lorentzian-distributed natural frequencies is treated separately by the Ott--Antonsen ansatz. They share the linear instability threshold but differ beyond linear order because thermal noise and quenched frequency heterogeneity are distinct forms of disorder. We then introduce the finite-load closure and the $\mathcal O(PN)$ evaluator, present the numerical strong-cue thresholds, and derive the fold and barrier asymptotics. The fitted slopes summarize the simulated sizes and are not asymptotic laws.

\section{Higher-order Kuramoto models}\label{sec:higherorderkuramoto}

Networks of weakly coupled limit-cycle oscillators provide a natural continuous-phase extension of Hopfield dynamics \cite{hopfield1982neural}. Their phases $\theta_i\in[0,2\pi)$ can encode information through relative phase. For identical oscillators with symmetric couplings in a co-rotating frame, the phase dynamics can be a gradient flow of a Lyapunov function; this equilibrium interpretation does not hold in general for heterogeneous natural frequencies. The classical weighted Kuramoto interaction is
\begin{equation}
\dot{\theta}_i
=\omega_i+\frac{1}{N}\sum_{j=1}^{N}J_{ij}\sin(\theta_j-\theta_i),
\end{equation}
where $\omega_i$ are the natural frequencies. It contains only the first Fourier harmonic of the pairwise phase difference. Although weighted Kuramoto networks can exhibit nontrivial multistability, this term alone does not provide the binary phase stabilization and enlarged Hebbian landscape sought here \cite{acebron2005kuramoto,aranson2002world}.

A minimal pairwise extension introduces a second Fourier harmonic,
\begin{equation}
    \dot{\theta}_i
    =
    \omega_i
    +
    \frac{1}{N}\sum_{j=1}^{N} J_{ij}\sin(\theta_j-\theta_i)
    +
    \frac{\lambda_2}{N}\sum_{j=1}^{N}\sin\!\bigl(2(\theta_j-\theta_i)\bigr),
    \label{eq:Kuramoto2}
\end{equation}
commonly termed the second-resonance Kuramoto model \cite{nishikawa2004oscillatory}. Here $\lambda_2$ is the intensive strength of the pairwise second-Fourier-harmonic interaction. This is the normalization used in Ref.~\cite{nishikawa2004oscillatory}, whose coefficient $\epsilon$ corresponds directly to $\lambda_2$. The term remains a two-body interaction; ``second harmonic'' refers to its Fourier dependence, not to its interaction order. The additional term creates two preferred phase differences near $0$ and $\pi$, stabilizing binary phase patterns and suppressing marginal drift.

The stored memories are independent Rademacher patterns
$\boldsymbol{\xi}^{\mu}=(\xi_1^\mu,\ldots,\xi_N^\mu)$,
$\mu=1,\ldots,P$, with
$\mathbb P(\xi_i^\mu=+1)=\mathbb P(\xi_i^\mu=-1)=1/2$.
We encode $\xi_i^\mu=+1$ by $\theta_i=0$ and $\xi_i^\mu=-1$ by $\theta_i=\pi$.
Retrieval of pattern $\mu$ is measured by the Mattis overlap
\[
z_\mu=\frac{1}{N}\sum_{j=1}^{N}\xi_j^\mu e^{i\theta_j}.
\]
The value $|z_\mu|=1$ represents the binary pattern up to a global phase. For even $p$ we use the overlap energies $F_2(z_\mu)=JN|z_\mu|^2/2$ and $F_p(z_\mu)=KN|z_\mu|^p/p!$ for $p>2$.
The corresponding even-$p$ Hamiltonian contains pairwise and higher-order cosine kernels,
\begin{eqnarray}
    H(\boldsymbol{\theta})
    &=&
    -\sum_{\mu = 1}^P
    \big(F_2(z_\mu)+ F_p(z_\mu)\big)
    \nonumber\\[2mm]
    &=&
    -\frac{J}{2N}\sum_{i,j=1}^N
    J_{ij}\cos(\theta_i-\theta_j)
    \nonumber\\
    -&&\!\!\!\!\!\!\!\frac{K}{N^{p-1}\,p!}
    \sum_{i_1,\cdots,i_p=1}^N
    K_{i_1\dots i_p}\,
    \cos(\Delta \theta_{1...p}),
    \label{eq:pKuramoto}
\end{eqnarray}
where $p$ is even, $
\Delta \theta_{1...p}\equiv
\theta_{i_1} + \cdots + \theta_{i_{p/2}}
- \theta_{i_{p/2+1}} - \cdots -\theta_{i_p},
$ and $ K_{i_1\!\dots i_p}\!=\!
\sum_{\mu=1}^{P} \xi_{i_1}^{\mu}\!\cdots\xi_{i_p}^{\mu}
$ is a Hebbian imprint of $P$ dense patterns
\cite{demircigil2017associative}.

We now specialize to $p=4$. The analysis comprises the single-pattern equilibrium landscape, a separate Ott--Antonsen reduction for frequency disorder, a finite-load crosstalk closure, and retrieval simulations with and without the pairwise channel. Hypergraph and simplicial representations can yield different higher-order dynamics, so the four-phase combination in Eq.~\eqref{eq:high-order-kuramoto} is part of the model definition \cite{zhang2023higher}.

We consider a population of $N$ phase oscillators whose dynamics combine
pairwise and four-body phase couplings (the $J$--$K$ Kuramoto model),

\begin{eqnarray}
\dot{\theta}_i
    &=&\; \omega_i+\frac{J}{N}\sum_{j=1}^{N}J_{ij}\, \sin\bigl(\theta_j-\theta_i\bigr)\nonumber \\
    &+&\frac{K}{6N^{3}}\sum_{j,k,\ell=1}^{N} K_{ijk\ell}\, \sin\bigl(\theta_j+\theta_k-\theta_\ell-\theta_i\bigr). \label{eq:high-order-kuramoto}
\end{eqnarray}
Here \(J\) is the overall strength of the pairwise interaction. Using the pattern ensemble defined above, the accompanying Hebbian matrix is
\begin{equation}
J_{ij}=\sum_{\mu=1}^{P}\xi_i^{\mu}\xi_j^{\mu} \quad(i\neq j),
\end{equation}
so that the first summation in Eq.~\eqref{eq:high-order-kuramoto} is the phase analogue of the Hopfield coupling. Similarly, \(K\) is the overall strength of the four-body interaction, with quartic Hebbian tensor
\begin{equation}
  K_{ijk\ell}
  =\sum_{\mu=1}^{P}
\xi_i^{\mu}\xi_j^{\mu}\xi_k^{\mu}\xi_\ell^{\mu},
    \label{kdef}
\end{equation}
symmetric under permutations of its indices. We use unrestricted sums in Eq.~\eqref{eq:high-order-kuramoto}, including coincident indices, because this convention gives the exact overlap factorization used below. Excluding coincident indices changes the intensive dynamics only through subleading finite-$N$ terms. The diagonal of $J_{ij}$ may be set to zero, since it contributes no phase torque. The factors \(1/N\) and \(1/N^{3}\) ensure that both interaction terms remain \(\mathcal{O}(1)\) as \(N\to\infty\). Throughout this paper, \(J_{ij}\) and \(K_{ijk\ell}\) denote pattern-derived couplings, while \(J\) and \(K\) control the overall interaction strengths.
The higher-order term is the phase analogue of the $p=4$ dense-Hopfield interaction \cite{krotov2016dense}.

For identical oscillators in the co-rotating frame ($\omega_i=\omega_0$), the deterministic drift is the gradient flow $\dot\theta_i=-\partial H/\partial\theta_i$ of the Lyapunov function $H$.

For heterogeneous natural frequencies, a stationary value of the Ott--Antonsen order-parameter modulus describes a time-independent macroscopic phase distribution in a frame rotating at $\omega_0$. It does not in general imply that every oscillator is frequency locked: for an unbounded Lorentzian distribution, partially synchronized stationary states can contain both locked and drifting oscillators. We therefore use the heterogeneous-frequency model only as a complementary synchronization calculation. The binary-memory interpretation and the retrieval simulations below refer to the identical-oscillator gradient/Langevin model. The existence and number of microscopic locked states for nonzero $\omega_i$ remain separate, nontrivial questions \cite{AGGV23}.

Related higher-order phase dynamics have been used in oscillator-based constraint solvers and in hardware emulations of higher-order Ising interactions \cite{mallick2022computational,bashar2023oscillator,nikhar2024all}. The all-to-all Hebbian tensor in Eq.~\eqref{eq:high-order-kuramoto} remains an additional implementation requirement.

\subsection{Equilibrium (static) mean-field theory}
\label{sec:equilibrium_mean_field}

We first consider the equilibrium mean-field theory in the retrieval frame of a single condensed pattern. The single-pattern equilibrium problem is the standard fully connected $XY$ mean-field model with combined quadratic and quartic overlap interactions; we reproduce its construction to establish the notation used for the Hebbian oscillator memory and for the subsequent fold and barrier calculations. Taking pattern $\mu=1$ as the retrieved memory, we perform the binary gauge rotation $\theta_i\mapsto \theta_i+\pi(1-\xi_i^1)/2$, so that $\xi_i^1=+1$ for all $i$. In the single-pattern, or leading low-load, approximation the Hebbian couplings are replaced by their uniform component, $J_{ij}\simeq1$ and $K_{ijk\ell}\simeq1$. The corresponding fully connected $XY$ Hamiltonian is
\begin{eqnarray}
H &=& -\frac{J}{2N}\sum_{i,j}\cos(\theta_j-\theta_i)\nonumber \\
&-&\;\frac{K}{24N^3}\sum_{i,j,k,\ell}\cos(\theta_k+\theta_\ell-\theta_i-\theta_j),
\end{eqnarray}
with order parameter $z=N^{-1}\sum_i e^{i\theta_i}$ and modulus $r=|z|\in[0,1]$. The Hamiltonian can be written as $H=-N U(r)$ with
\begin{equation}
U(r)=\frac{J}{2}r^2+\frac{K}{24}r^4,
\qquad
U'(r)=Jr+\frac{K}{6}r^3.
\label{eq:single-pattern-energy-density}
\end{equation}

We use the equilibrium Langevin version of Eq.~\eqref{eq:high-order-kuramoto}. The oscillators are identical, $\omega_i=\omega_0$, and after moving to the co-rotating frame we add independent Gaussian white noise,
\begin{equation}
d\theta_i=-\frac{\partial H}{\partial\theta_i}\,dt+\sqrt{2/\beta}\,dW_i(t).
\label{eq:SDE_definition}
\end{equation}
This reversible dynamics has stationary measure proportional to $\exp(-\beta H)$. The exact leading-$N$ constrained free-energy, or large-deviation rate function, for the physical magnetization modulus is
\begin{equation}
\Phi(r)
=
 a(r)r-\ln I_0\!\bigl(a(r)\bigr)-\beta U(r),
\qquad
r=\frac{I_1(a(r))}{I_0(a(r))},
\label{eq:constrained-free-energy}
\end{equation}
where $a(r)\ge0$ is the conjugate field and $\beta=1/T$ with $k_B=1$. Additive constants independent of $r$ are omitted.

For locating stationary branches it is convenient to use the auxiliary variational potential
\begin{equation}
F_{\rm aux}(r)
=
\frac{\beta J}{2}r^2+\frac{\beta K}{8}r^4
-\ln I_0\!\left[\beta\left(Jr+\frac{K}{6}r^3\right)\right].
\label{F}
\end{equation}
Writing $g(X)=I_1(X)/I_0(X)$ and $X(r)=\beta U'(r)$, its derivative factorizes as
\begin{equation}
\begin{aligned}
F_{\rm aux}'(r)
&=
\beta U''(r)\,[r-g(X(r))]\\
&=
\beta\left(J+\frac{K}{2}r^2\right)[r-g(X(r))].
\label{eq:Faux-derivative}
\end{aligned}
\end{equation}
For $J>0$ and $K\ge0$, the stationary points therefore satisfy
\begin{equation}
 r
 =
 \frac{I_1\!\left[\beta\left(Jr+\frac{K}{6}r^3\right)\right]}
 {I_0\!\left[\beta\left(Jr+\frac{K}{6}r^3\right)\right]}.
\label{eq:self-consistency}
\end{equation}
At every stationary point $r_*$, the conjugate field obeys $a(r_*)=\beta U'(r_*)$, and hence
\begin{equation}
F_{\rm aux}(r_*)=\Phi(r_*).
\label{eq:stationary-potential-equality}
\end{equation}
Thus $F_{\rm aux}$ gives the correct stationary branches, coexistence condition, and free-energy differences between extrema. It is not, however, identical to the constrained profile $\Phi(r)$ away from stationary points; in particular, its off-shell curvature is an auxiliary diagnostic rather than the physical order-parameter fluctuation curvature.

\paragraph{Landau expansion and the critical and tricritical points.}
The entropy part of Eq.~\eqref{eq:constrained-free-energy} has the expansion
\begin{equation}
a(r)r-\ln I_0(a(r))
=r^2+\frac14 r^4+\frac{5}{36}r^6+O(r^8).
\end{equation}
Consequently, the exact dimensionless constrained free energy has the Landau form
\begin{equation}
\Phi(r)=b_2r^2+b_4r^4+b_6r^6+O(r^8),
\label{ff}
\end{equation}
with
\begin{equation}
 b_2=1-\frac{\beta J}{2},
 \qquad
 b_4=\frac14-\frac{\beta K}{24},
 \qquad
 b_6=\frac{5}{36}>0.
\label{coeffs}
\end{equation}
The coefficient $b_2$ changes sign on the incoherent spinodal $x_{\mathrm P}=\beta J=2$. On that line,
\begin{equation}
 b_4\big|_{x=2}
 =\frac{3-C}{12},
 \qquad C:=K/J.
\end{equation}
For $C<3$, $b_4>0$ and the nonzero minimum emerges continuously, with $r\propto(T_c-T)^{1/2}$. At $C=3$, the quartic Landau coefficient vanishes while $b_6=5/36$ remains positive, giving the thermodynamic tricritical point $(x,C)=(2,3)$. For $C>3$, the small-$r$ branch is subcritical and a nonzero minimum can coexist with the incoherent minimum for $x<2$.

\begin{figure}[!ht]
    \centering
    \includegraphics[width=1\linewidth]{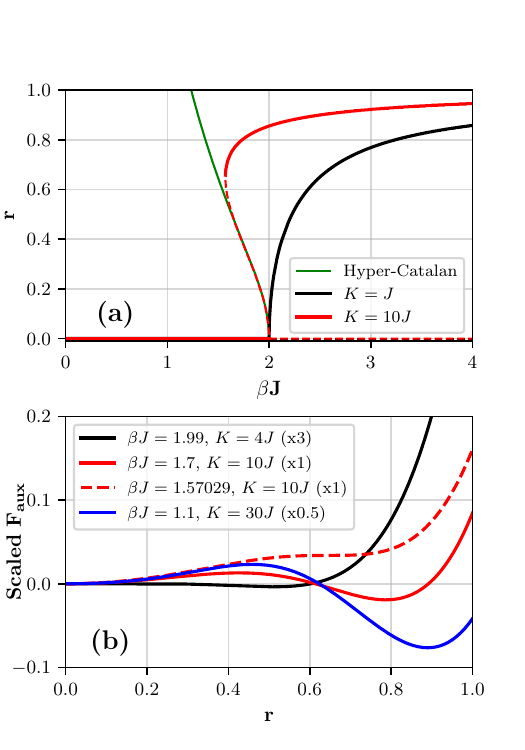}
    \vspace{-1em}
    \caption{(a) Equilibrium order-parameter branches of the single-pattern model. Solid curves plot stationary solutions of $r=g\!\bigl(X(x,r;C)\bigr)$ versus $x=\beta J$ for $C=K/J=1$ (black) and $C=10$ (red); solid and dashed portions denote locally stable and unstable radial extrema, respectively. The green curve is the truncated hyper-Catalan approximation in Eq.~\eqref{eq:rHC_stream_new}. Its agreement is visual and limited to the plotted range away from the fold; no uniform error bound is claimed. At $x=2$ the $C=1$ model has a continuous onset, whereas $C=10$ has a subcritical branch and a bistable interval. (b) Dimensionless auxiliary-potential profiles $F_{\rm aux}(r)$ for $(C,x)=\{(4,1.99),(10,1.70),(10,1.57029),(30,1.10)\}$. Multiplicative plotting factors are indicated in the legend. For $(10,1.70)$, the unstable radial extremum is at $r_u\simeq0.43$ and the memory minimum at $r_s\simeq0.79$; at the fold $(10,1.57029)$ they coalesce near $r\simeq0.65$. For $C=30$, the nonzero extrema lie outside the quantitative range of the small-$r$ sextic expansion, so the full non-polynomial mean-field functions are required.}
\label{fig:oscmemory}
\end{figure}

\begin{figure}[t]
\centering
\includegraphics[width=1\linewidth]{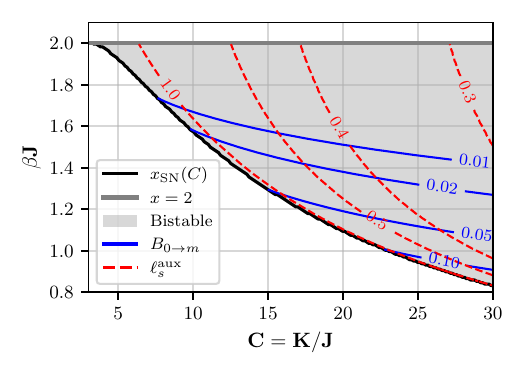}
\caption{Equilibrium landscape of the single-pattern model in the $(C,x)$ plane, with $C=K/J>3$ and $x=\beta J$. Solid blue contours show the incoherent-side barrier $B_{0\to m}=\Phi(r_u)-\Phi(0)$; by Eq.~\eqref{eq:stationary-potential-equality}, these values are identical to $F_{\rm aux}(r_u)-F_{\rm aux}(0)$. They do not represent the memory-retention barrier $B_{m\to0}=\Phi(r_u)-\Phi(r_s)$. Dashed red contours show the auxiliary inverse-curvature diagnostic $\ell_s^{\rm aux}=[F_{\rm aux}''(r_s)]^{-1/2}$ at the memory minimum. It is neither the geometric basin width nor the physical finite-$N$ radial fluctuation scale, which is controlled by $\Phi''(r_s)$. The shaded strip $x_{\mathrm{SN}}(C)<x<2$ is bistable: the lower boundary is the fold where $r_s$ and $r_u$ merge, and the upper boundary is the incoherent spinodal. The two directional barriers have opposite limiting behavior across this strip: $B_{m\to0}\to0$ at the fold, whereas $B_{0\to m}\to0$ as $x\to2^-$.}
\label{fig:barriers}
\end{figure}

\paragraph{Thermodynamic first-order branch.}
For $C>3$, the constrained free-energy profile develops a nonzero minimum and an intervening maximum. The exact sextic Landau coefficient is positive, but the small-$r$ truncation is quantitatively controlled only near the tricritical point; for ratios such as $C=10$ and $C=30$, the full constrained function in Eq.~\eqref{eq:constrained-free-energy}, or equivalently the stationary values of Eq.~\eqref{F}, must be used. The equal-depth transition occurs at $x_{\mathrm{coex}}(C)<2$. Since $x=J/T$, this means that on cooling the discontinuous transition occurs at $T_{\mathrm{coex}}>J/2$; the point $x=2$ is the lower-temperature spinodal at which the incoherent state loses local stability. The thermodynamic tricritical point belongs only to the equilibrium Langevin model and must not be identified with the nonlinear onset of the heterogeneous-frequency Ott--Antonsen model discussed next.

Within the equilibrium model, $C<3$ gives a continuous onset, $C=3$ is tricritical, and $C>3$ produces a bistable interval and a discontinuous transition. These statements concern the single-condensed-pattern landscape; they do not by themselves establish a capacity law at finite load. The fold and barrier asymptotics are derived in \hyperref[sec:equilibrium-details]{Analytical details of equilibrium bistability}.

\subsection{Dynamical mean-field approximation for heterogeneous frequencies}
We now turn to the distinct deterministic model with Lorentzian natural-frequency distribution
\begin{equation}
 g(\omega)=\frac{\Delta}{\pi[(\omega-\omega_0)^2+\Delta^2]},
\end{equation}
and no Langevin noise. The Ott--Antonsen ansatz gives $\dot\psi=\omega_0$ and the closed amplitude equation \cite{ott2008low,skardal2020higher}
\begin{equation}\label{eq:SA-dynamics}
\dot r
=
-\Delta r
+\frac{J}{2}r(1-r^2)
+\frac{K}{12}r^3(1-r^2).
\end{equation}
For $K>0$, the nonzero stationary moduli are
\begin{equation}\label{eq:OA-roots}
 r_{\pm}
 =
 \left[
 \frac{K-6J\pm\sqrt{(6J+K)^2-48\Delta K}}{2K}
 \right]^{1/2},
\end{equation}
provided that the expression in square brackets lies in the physical interval $[0,1]$. Because $r=|z|\ge0$, there is no additional outer $\pm$ sign. In the pairwise limit $K=0$, the nonzero stationary branch is $r^2=1-2\Delta/J$ when $J>2\Delta$. A stationary $r>0$ describes a stationary macroscopic phase distribution in the frame rotating at $\omega_0$; it need not imply microscopic locking of every oscillator.

Expanding Eq.~\eqref{eq:SA-dynamics} gives
\begin{equation}
 \dot r
 =r\left[
 \frac{J}{2}-\Delta
 +\left(\frac{K}{12}-\frac{J}{2}\right)r^2
 -\frac{K}{12}r^4
 \right].
 \label{eq:OA-amplitude-expanded}
\end{equation}
The incoherent state changes linear stability at $\Delta_c=J/2$. On this line the cubic coefficient changes sign at
\begin{equation}
 K=6J.
\end{equation}
Consequently, the OA onset is supercritical for $K<6J$ and subcritical for $K>6J$, with a codimension-two point at $(\Delta/J,K/J)=(1/2,6)$. This differs from the equilibrium thermodynamic tricritical ratio $K/J=3$. The difference is expected because quenched frequency disorder and annealed thermal noise generate different nonlinear amplitude equations.

Setting
\begin{equation}\label{eq:matching}
 \Delta=\frac{1}{\beta}
\end{equation}
merely matches the two \emph{linear} instability thresholds, $\Delta=J/2$ and $\beta J=2$. It does not identify the nonlinear dynamics, the stationary branches, or the equilibrium free-energy landscape. In particular, equilibrium curvatures are not used as OA relaxation rates. For an actual stable nonzero OA fixed point $r_*$, direct linearization of Eq.~\eqref{eq:SA-dynamics} gives
\begin{equation}
 \lambda_{\mathrm{OA}}
 =-\left.\frac{d\dot r}{dr}\right|_{r_*}
 =2r_*^2\left(\frac{J}{2}-\frac{K}{12}+\frac{K}{6}r_*^2\right)>0.
 \label{eq:OA-rate}
\end{equation}
We use the Lorentzian distribution because it admits this exact low-dimensional closure; for more general frequency distributions and phase lags, synchronization transitions can be nonuniversal \cite{omelchenko2012nonuniversal}. Higher-order interactions in neuronal oscillator models likewise produce distinct nonequilibrium synchronization phenomena \cite{parastesh2022synchronization}.

\subsection*{Mean-field signal--to--noise closure at finite load}

To describe retrieval at finite load, we use a phenomenological cavity/signal--to--noise closure rather than a rigorous finite-$N$ fluctuation expansion. A controlled mesoscopic fluctuation theory would require methods of the type developed in Refs.~\cite{Buendia2025,omel2026mean} and is not attempted here. 
Writing Eq.~\eqref{eq:high-order-kuramoto} in terms of the complex overlaps
\begin{equation}
m_\mu
=
r_\mu e^{i\phi_\mu}
=
\frac{1}{N}\sum_{j=1}^{N}\xi_j^\mu e^{i\theta_j}
\label{eq:complex-overlap-finite-load}
\end{equation}
gives
\begin{equation}
\dot\theta_i
=
\omega_i
+
\sum_{\mu=1}^{P}
\left(
Jr_\mu+\frac{K}{6}r_\mu^{3}
\right)
\xi_i^\mu
\sin(\phi_\mu-\theta_i).
\label{eq:dynamicsOP}
\end{equation}
We condition on a retrieval state in which pattern $\mu=1$ is condensed. After the associated binary gauge rotation, $\xi_i^1\equiv1$, and we write $r\equiv r_1$ and $\phi\equiv\phi_1$. The remaining overlaps are assumed to be non-condensed, $r_\mu=\mathcal O(N^{-1/2})$. For fixed $P$ and large $N$, each non-condensed overlap acts as an $\mathcal O(N^{-1/2})$ perturbation and the retrieval branch persists by continuity of the attracting fixed point; the closure below addresses instead the finite-load regime $P=\alpha N$, where the aggregate crosstalk is not perturbative.

The closure uses three explicit assumptions. First, removing one non-condensed pattern changes the macroscopic retrieval state only by $o(1)$ (a cavity assumption). Second, in that cavity state, the phases are independent of the removed pattern bits to leading order. Third, the aggregate contributions of the $P-1$ non-condensed patterns obey a weak-dependence/Lindeberg condition. These assumptions are standard in signal--to--noise treatments of associative memories, but they are not consequences of Eq.~\eqref{eq:high-order-kuramoto} proved here.

We distinguish the complex crosstalk field from the scalar torque that it induces. The non-condensed complex field at site $i$ is
\begin{equation}
C_i
:=
\sum_{\mu=2}^{P}
\left(
Jm_\mu+\frac{K}{6}m_\mu^2\bar m_\mu
\right)
\xi_i^\mu,
\label{eq:complex-crosstalk-field}
\end{equation}
and its projection into the phase equation is
\begin{equation}
\eta_i(\theta_i)
:=
\Im\!\left(e^{-i\theta_i}C_i\right).
\label{eq:projected-crosstalk-torque}
\end{equation}
Thus
\begin{equation}
\dot\theta_i
=
\omega_i
+
\left(Jr+\frac{K}{6}r^{3}\right)
\sin(\phi-\theta_i)
+
\eta_i(\theta_i).
\label{eq:dynamicsGN}
\end{equation}

For the pairwise part,
\begin{equation}
C_i^{(J)}=J\sum_{\mu=2}^{P}\xi_i^\mu m_\mu,
\label{eq:pairwise-complex-crosstalk}
\end{equation}
and the exact leave-one-site decomposition
\begin{equation}
m_\mu
=
m_\mu^{(i)}
+
\frac{1}{N}\xi_i^\mu e^{i\theta_i},
\qquad
m_\mu^{(i)}
:=
\frac{1}{N}\sum_{j\neq i}\xi_j^\mu e^{i\theta_j}
\label{eq:loo-overlap-binary}
\end{equation}
yields
\begin{align}
C_i^{(J)}
&=
J\sum_{\mu=2}^{P}\xi_i^\mu m_\mu^{(i)}
+
J\frac{P-1}{N}e^{i\theta_i}
\nonumber\\
&=:C_i^{(J),\circ}+J\frac{P-1}{N}e^{i\theta_i}.
\label{eq:crosstalk-self-split}
\end{align}
The explicit self-term produces no torque because
\begin{equation}
\Im\!\left(e^{-i\theta_i}C_i^{(J)}\right)
=
\Im\!\left(e^{-i\theta_i}C_i^{(J),\circ}\right).
\label{eq:self-term-no-torque}
\end{equation}
Under the cavity independence assumption, $m_\mu^{(i)}$ is independent of $\xi_i^\mu$ to leading order and hence
\begin{equation}
\mathbb E_{\rm cav}\!\left[C_i^{(J),\circ}\right]=0.
\label{eq:centered-crosstalk-mean-zero}
\end{equation}
This is an approximate cavity statement, not a conditional expectation at fixed phases of the full evolved system.

The same expansion shows that the quartic torque has only a subleading cavity mean. At fixed cavity phases,
\begin{align}
&\mathbb E_{\rm cav}\!\left[
\xi_i^{\mu}\,
\Im\!\left(e^{-i\theta_i}m_\mu^{2}\bar m_\mu\right)
\right]\nonumber\\
&\qquad=
\frac{1}{N^{3}}\sum_{j\neq i}\sin\!\left[2(\theta_j-\theta_i)\right]
=
\mathcal O(N^{-2}).
\label{eq:quartic-cavity-mean}
\end{align}
Thus, summing $P-1=\alpha N+\mathcal O(1)$ non-condensed patterns gives a mean quartic torque of order $\mathcal O(\alpha K/N)$. In particular, the positive mean of the moduli $r_\mu$ does not by itself imply a drift of order $\alpha J\sqrt N$: the torque depends on signed complex overlaps and on their phases. We henceforth write $C_i^\circ:=C_i-\mathbb E_{\rm cav}C_i$ for the full centred complex field, including both channels.

The individual pattern contributions are non-Gaussian and share the same dynamical phase configuration. We therefore invoke a Gaussian closure only under the cavity and weak-dependence assumptions stated above.

The moments of a non-condensed complex overlap need not be circular. Introduce the second angular moment
\begin{equation}
q_2:=\frac{1}{N}\sum_{j=1}^{N}e^{2i\theta_j}.
\label{eq:q2}
\end{equation}
The leading cavity Gaussian moments are
\begin{align}
\mathbb E_{\rm cav}|m_\mu|^2
&=\frac{1}{N}+o(N^{-1}),\nonumber\\
\mathbb E_{\rm cav}|m_\mu|^4
&=\frac{2+|q_2|^2}{N^2}+o(N^{-2}),\nonumber\\
\mathbb E_{\rm cav}|m_\mu|^6
&=\frac{6+9|q_2|^2}{N^3}+o(N^{-3}).
\label{eq:noncondensed-moments}
\end{align}
The coefficients $2$ and $6$ used for a circular complex Gaussian are recovered when $q_2\simeq0$; near a binary retrieved state, $q_2\simeq1$, and the corresponding coefficients are $3$ and $15$.

Consequently, the total second moment of the complex crosstalk field is estimated as
\begin{align}
V_C
&:=\mathbb E_{\rm cav}|C_i^\circ|^2\nonumber\\
&\simeq
\alpha\!\left[
J^2
+
\frac{JK}{3N}\left(2+|q_2|^2\right)
+
\frac{K^2}{36N^2}\left(6+9|q_2|^2\right)
\right].
\label{eq:sigma2}
\end{align}
For fixed $J>0$, the pairwise crosstalk reaches order unity on the scale $P=\mathcal O(N)$. When $J=0$, formally extrapolating the quartic contribution gives an order-one scale $P=\mathcal O(N^3)$; this extrapolation lies outside the fixed-$\alpha$ regime in which the cavity and weak-dependence assumptions are formulated and is not controlled. These are signal--to--noise crossover scales, not capacity theorems; in particular, Eq.~\eqref{eq:sigma2} does not derive the finite-size exponents measured below.

\begin{figure}[!ht]
    \centering
    \includegraphics[width=1\linewidth]{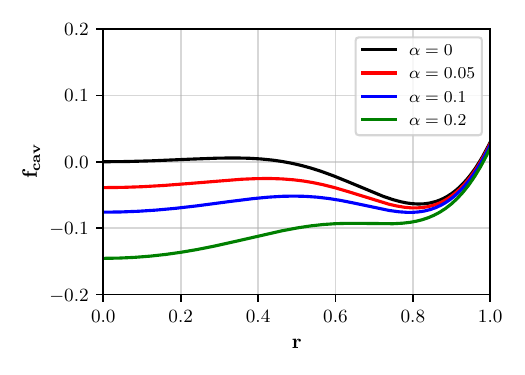}
    \vspace{-1.5em}
\caption{Illustrative scalar signal--to--noise potential $f_{\rm cav}(r)$ from Eq.~\eqref{eq:freeEnergyNoise}, plotted for increasing load $\alpha$ at fixed $\beta J=1.79$ and $K=10J$. The curves use the circular-moment convention $q_2=0$ and the leading large-$N$ approximation $\sigma_\parallel^2=V_C\simeq\alpha J^2$, with the $O(N^{-1})$ and $O(N^{-2})$ terms in Eq.~\eqref{eq:sigma2} omitted. Within this phenomenological closure, increasing crosstalk flattens the retrieval well. The plot should not be interpreted as a rigorous finite-load free-energy calculation.}
    \label{fig:free-energy-gaussian-crosstalk}
\end{figure}

The effective complex field is written
\begin{equation}
H_i^{\mathrm{eff}}
=
h+C_i^\circ,
\qquad
h:=Jr+\frac{K}{6}r^3.
\label{eq:effective-complex-field}
\end{equation}
For the scalar closure used in Fig.~\ref{fig:free-energy-gaussian-crosstalk}, we keep only the longitudinal component of $C_i^\circ$ along the retrieval direction and set
\begin{equation}
\eta_\parallel\sim\mathcal N(0,\sigma^2),
\qquad
\sigma^2\equiv\sigma_\parallel^2\simeq V_C.
\label{eq:longitudinal-variance}
\end{equation}
This convention is phenomenological. An isotropic complex field would instead have longitudinal variance $V_C/2$, whereas a nearly binary retrieval state can be strongly anisotropic. For the illustrative curves below we use the circular-moment choice $q_2=0$ and the convention $\sigma_\parallel^2=V_C$.

Holding this effective variance fixed, the one-site scalar closure has field $H=h+\sigma z$, with $z\sim\mathcal N(0,1)$, and the effective potential
\begin{equation}
f_{\rm cav}(r)
=
\frac{J}{2}r^{2}
+
\frac{K}{8}r^{4}
-
\frac{1}{\beta}
\int Dz\,
\ln I_{0}\!\left[\beta\left(h+\sigma z\right)\right],
\label{eq:freeEnergyNoise}
\end{equation}
where
\begin{equation}
Dz=(2\pi)^{-1/2}e^{-z^{2}/2}\,dz.
\end{equation}
Its stationarity condition is
\begin{equation}
r
=\int\!Dz\,
\frac{I_{1}[\beta(h+\sigma z)]}{I_{0}[\beta(h+\sigma z)]},
\qquad
\frac{\partial f_{\rm cav}}{\partial r}=0,
\label{eq:selfconsistencynoise}
\end{equation}
provided that $\sigma$ is treated as an external closure parameter. We use this construction only for a qualitative comparison of load-induced flattening; it is not an exact quenched free energy of the finite-load dynamics.

In the zero-temperature limit, $\ln I_0(\beta x)\simeq\beta|x|$, and the same scalar closure gives
\begin{align}
f_{{\rm cav},T=0}(r)
=\frac{J}{2}r^{2}+\frac{K}{8}r^{4}
-\Bigl[&\sigma\sqrt{\tfrac{2}{\pi}}\,e^{-h^{2}/(2\sigma^{2})}\nonumber\\
&+h\,\operatorname{erf}\!\left(\frac{h}{\sqrt{2}\sigma}\right)\Bigr].
\label{eq:zeroT-signal-noise}
\end{align}
The disorder term is not multiplied by an additional factor of $J$. Thus it remains nonzero in the pure-quartic case $J=0$, for which Eq.~\eqref{eq:sigma2} gives $\sigma^2=\mathcal O(PK^2/N^3)$.

\section{Numerical simulations for memory retrieval}\label{sec:numerics}
We first compare the low-load bifurcation with equilibrium theory, then examine a finite-load trajectory and the strong-cue recovery and retention thresholds.

\subsection*{Efficient evaluation of the four-body term}

To simulate Eq.~\eqref{eq:high-order-kuramoto} efficiently, we avoid explicitly forming the full four-body Hebbian tensor $K_{ijk\ell}$, which would require $\mathcal{O}(N^4)$ memory and computation. Instead, we exploit the binary nature of the stored patterns $\xi_i^{\mu} \in \{\pm 1\}$ and rewrite the interaction terms using global pattern overlaps.

Define the unnormalized overlaps
\begin{equation}
    S_+^{\mu} = \sum_j \xi_j^{\mu} e^{i\theta_j}, \qquad S_-^{\mu} = \overline{S_+^{\mu}}.
\end{equation}
The four-body contribution to $\dot\theta_i$ is
\begin{equation}
    \frac{K}{6N^3} \sum_{\mu=1}^P \xi_i^{\mu}\,
    \mathrm{Im} \left[(S_+^{\mu})^2 S_-^{\mu} e^{-i\theta_i}\right].
\end{equation}

The resulting update costs $\mathcal O(PN)$ operations per time step and avoids storing an $N^4$ tensor.

Finite-temperature simulations integrate the Langevin form of Eq.~\eqref{eq:high-order-kuramoto},
\[
d\theta_i=F_i(\boldsymbol\theta)\,dt+\sqrt{2/\beta}\,dW_i(t),
\]
where $F_i$ is the deterministic drift. We use Euler--Maruyama for this SDE. Figure~\ref{fig:ScalingPlot_zeroT} instead uses identical oscillators with $\omega_i=0$, no dynamical noise, and explicit Euler integration. The Lorentzian-frequency model is used only in the Ott--Antonsen calculation; $\Delta$ and $T=1/\beta$ are compared only through the linear thresholds.

Unless stated otherwise, a structured cue is perturbed once at $t=0$ according to
\begin{equation}
\begin{aligned}
\theta_i(0)&=\theta_i^{\rm cue}+\eta_i,\\
\eta_i&\stackrel{\rm iid}{\sim}{\rm Unif}[-\delta_{\rm init},\delta_{\rm init}],
\qquad \delta_{\rm init}=0.1\ \text{rad}.
\end{aligned}
\label{eq:initial-jitter}
\end{equation}
This perturbation is distinct from the Langevin noise and is not reapplied. In the deterministic zero-temperature runs it moves the cue off the exactly binary invariant manifold, on which every sine torque vanishes. In the target gauge, a binary cue with exact pre-jitter overlap $\epsilon$ has
\begin{equation}
\mathbb E\!\left[z_\mu(0)\mid\epsilon\right]
=\epsilon\,\operatorname{sinc}(\delta_{\rm init}),
\qquad
\operatorname{sinc}(\delta):=\frac{\sin\delta}{\delta}.
\label{eq:jitter-overlap}
\end{equation}
For $\delta_{\rm init}=0.1$, $\operatorname{sinc}(0.1)=0.998334\ldots$; exact pre-jitter overlaps $1$, $0.9$, and $0.95$ therefore give expected complex overlaps $0.9983$, $0.8985$, and $0.9484$.

\subsection*{Low- and finite-load comparisons}
We compare the equilibrium single-pattern prediction with simulations at $N=500$ and $P=1$. This removes pattern crosstalk and isolates the low-load bifurcation, although finite-size fluctuations and the finite parameter-sweep time can still shift a discontinuous jump.

\begin{figure}[!ht]
    \centering
    \includegraphics[width=1\linewidth]{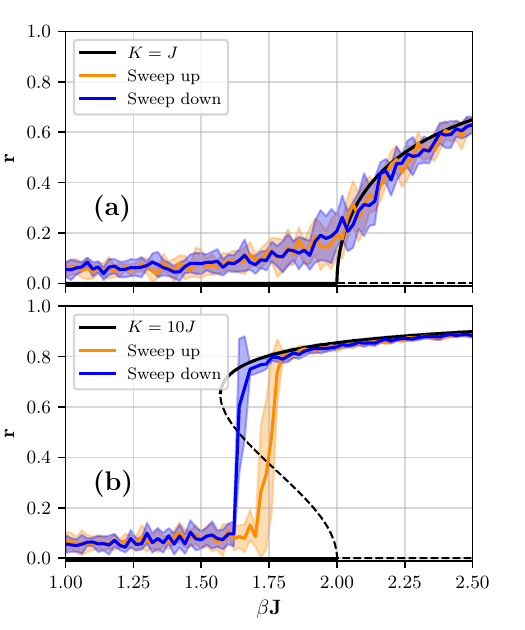}
    \vspace{-1.5em}
    \caption{Upward (blue) and downward (orange) continuation sweeps of $\beta J$ for $N=500$ and one stored pattern ($P=1$), with $J=1$ and $C=K/J=1$ (top) or $10$ (bottom). For each sweep direction, $\beta J$ is varied over $[1,2.5]$ in increments of $\Delta(\beta J)=0.01$. At every parameter value the Langevin equation is integrated by Euler--Maruyama with $\Delta t=0.01$ for $T_{\rm sim,step}=300$; the overlap is recorded at the end of that segment and the terminal phase configuration initializes the next parameter value. The first point of each realization is initialized at the stored target pattern and perturbed by the one-time angular jitter in Eq.~\eqref{eq:initial-jitter}, giving expected initial complex overlap $\operatorname{sinc}(0.1)\simeq0.9983$. The upward and downward sweeps are generated as independent ensembles. For each direction, $N_{\mathrm{trial}}=20$ realizations use independent initial jitters and Langevin-noise paths. Within a realization, the terminal state at one value of $\beta J$ initializes the next. Solid curves show the ensemble mean of $r$, and shaded bands show one standard deviation across the $20$ realizations. Black curves are equilibrium mean-field branches from Eq.~\eqref{eq:self-consistency}. Panel (a) is consistent with continuous onset. Panel (b) shows qualitative hysteresis, while the jump locations are displaced from the thermodynamic branches.}
    \label{fig:hysteresis_low}
\end{figure}

Figure~\ref{fig:hysteresis_low} shows the qualitative change as $K/J$ increases. For $K/J<3$, the onset is continuous, whereas the $K/J=10$ sweep exhibits a discontinuous jump and hysteresis, consistent with equilibrium bistability. Panel~(b) does not reproduce the thermodynamic jump locations quantitatively. Finite size and the finite continuation time are plausible contributors, but the present data do not establish how the displacement scales with $N$.

\begin{figure}[!ht]
    \centering
    \includegraphics[width=1\linewidth]{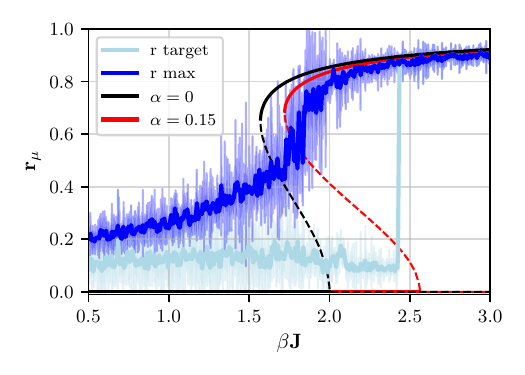}
    \vspace{-1.5em}
    \caption{Single downward continuation sweep of $\beta J$ for memory load $\alpha=0.15$, with $J=1$, $C=K/J=10$, $N=100$, and $P=15$. The sweep starts at $\beta J=3.0$ from the target pattern, perturbed by the one-time angular jitter in Eq.~\eqref{eq:initial-jitter}, so that its expected initial complex overlap is $\operatorname{sinc}(0.1)\simeq0.9983$. It then proceeds to $\beta J=0.5$ in decrements of $\Delta(\beta J)=0.01$. At every parameter value the Langevin equation is integrated by Euler--Maruyama with $\Delta t=0.01$ for $T_{\rm sim,step}=300$, and the terminal state initializes the next point. The light-blue curve shows the overlap with the target pattern, and the dark-blue curve the largest overlap over all stored patterns. The values are averaged over each integration segment and then smoothed along the sweep with a one-dimensional Gaussian filter of width $\sigma=3$ sweep points. At each point the shaded region spans the raw and smoothed values. Mean-field predictions without crosstalk (black, Eq.~\eqref{eq:self-consistency}) and with the phenomenological finite-load scalar closure (red, Eq.~\eqref{eq:selfconsistencynoise}, using the same $q_2=0$ and $\sigma_\parallel^2=V_C$ convention as Fig.~\ref{fig:free-energy-gaussian-crosstalk}) are shown for comparison. The run contains both thermal fluctuations and pattern crosstalk; the comparison is therefore qualitative and does not isolate the two effects.}
    \label{fig:hysteresis_high}
\end{figure}

We next consider a finite pattern load $\alpha=P/N$. Increasing $P$ at fixed $N$ introduces quenched interference and competing overlaps. Figure~\ref{fig:hysteresis_high} compares one microscopic run at $\alpha=0.15$ with the single-pattern prediction and with the phenomenological scalar cavity closure. The comparison is illustrative rather than a controlled validation of the closure.

We plot the static single-pattern prediction without crosstalk (black) and the scalar cavity closure with crosstalk (red). In this illustrative calculation, including the load shifts the predicted stability boundary, with the largest change near the fold. In the microscopic sweep, the target pattern (light blue) is initially retrieved and the trajectory later switches to a competing stored pattern (dark blue) as $\beta J$ decreases. Because the run contains both Langevin noise and quenched pattern interference, the early transition cannot be attributed to crosstalk alone. The red curve is qualitatively closer to the observed transition than the noiseless curve, but the comparison is based on a single finite-size trajectory and does not independently validate the Gaussian closure.

\subsection*{Finite-size strong-cue recovery and retention thresholds}

For the finite-size threshold data, we fix $N$, vary the number of stored patterns $P$, and measure the terminal overlap with a randomly selected target. Each trial consists of (i) generating $P$ independent Rademacher patterns; (ii) selecting one target pattern; (iii) constructing a binary cue by shifting $n_{\rm flip}$ randomly selected target phases by $\pi$; (iv) applying the one-time angular jitter in Eq.~\eqref{eq:initial-jitter}; and (v) integrating the appropriate finite- or zero-temperature dynamics to the prescribed final time $t_f=T_{\rm sim}$. The exact finite-$N$ pre-jitter overlap of the binary cue is
\begin{equation}
\epsilon_N=1-\frac{2n_{\rm flip}}{N},
\label{eq:finiteN-cue-overlap}
\end{equation}
where the integer $n_{\rm flip}$ is chosen by unbiased stochastic rounding. Write
$n_{\rm flip}^{*}=N(1-\epsilon_0)/2=m+\rho$,
where $m=\lfloor n_{\rm flip}^{*}\rfloor$ and $\rho\in[0,1)$. We take $n_{\rm flip}=m$ with probability $1-\rho$ and $n_{\rm flip}=m+1$ with probability $\rho$. Hence $\mathbb E[\epsilon_N]=\epsilon_0$ and $|\epsilon_N-\epsilon_0|<2/N$. The independent jitter in Eq.~\eqref{eq:initial-jitter} is then applied once. Conditional on $\epsilon_N$, the expected complex overlap is $\epsilon_N\operatorname{sinc}(0.1)$. Finite-temperature trials use independent Langevin-noise realizations; zero-temperature trials have no dynamical noise. At nonzero temperature the terminal value is a sample from a stationary or metastable retrieval regime, not a phase-locked fixed point.

\begin{figure}[!ht]
    \centering
    \includegraphics[width=0.95\linewidth]{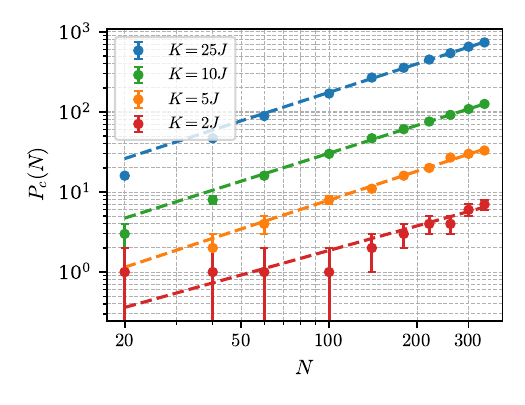}
    \vspace{-1.5em}
    \caption{Finite-temperature strong-cue recovery/retention threshold of the $J$--$K$ Kuramoto model. Each trial begins from a binary cue with requested pre-jitter target overlap $\epsilon_0=0.9$ and exact finite-$N$ overlap $\epsilon_N$ from Eq.~\eqref{eq:finiteN-cue-overlap}, then receives the one-time perturbation $\eta_i\sim{\rm Unif}[-0.1,0.1]$ in Eq.~\eqref{eq:initial-jitter}; for a cue with $\epsilon_N=0.9$, the expected post-jitter complex overlap is approximately $0.8985$. The cue evolves under the Langevin form of Eq.~\eqref{eq:high-order-kuramoto}, integrated by Euler--Maruyama with $\Delta t=10^{-3}$ to $T_{\rm sim}=150$, and the final time step is recorded. For $N_{\rm trial}=100$, $P_c(N)$ is estimated by the integer binary-search protocol described in the text as the largest passing load for which the mean terminal overlap satisfies $\overline r_f\ge0.85$. Because this terminal threshold lies below the initial overlap, the measurement tests finite-time preservation of a strong cue rather than completion from a weaker cue. We set $J=1$, $\beta J=4$, and use $C=K/J=2$ (red), $5$ (orange), $10$ (green), and $25$ (blue). The brackets reflect the final binary-search resolution rather than statistical confidence intervals. Nominal fits $P_c\propto N^{\gamma_{\rm eff}}$ over the displayed range give $\gamma_{\rm eff}=1.02$, $1.20$, $1.16$, and $1.18$, respectively; these are descriptive finite-size slopes, not asymptotic exponents.}
    \label{fig:ScalingPlot_finiteT}
\end{figure}

\begin{figure}[!ht]
    \centering
    \includegraphics[width=0.95\linewidth]{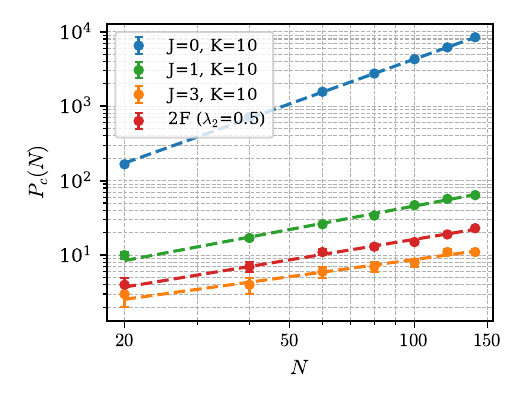}
    \vspace{-1.5em}
    \caption{Zero-temperature finite-time strong-cue recovery/retention threshold for identical oscillators with $\omega_i=0$. Each trial begins from a corrupted binary target cue with requested pre-jitter overlap $\epsilon_0=0.95$ and exact finite-$N$ overlap $\epsilon_N$ given by Eq.~\eqref{eq:finiteN-cue-overlap}, followed by the one-time independent perturbation $\eta_i\sim{\rm Unif}[-0.1,0.1]$ in Eq.~\eqref{eq:initial-jitter}. For a cue with $\epsilon_N=0.95$, the expected post-jitter complex overlap is $0.95\operatorname{sinc}(0.1)\simeq0.9484$. No dynamical noise is added. The equations are integrated by explicit Euler with $\Delta t=10^{-3}$ to $T_{\rm sim}=150$, and the final time step is recorded. For $N_{\rm trial}=100$, $P_c(N)$ is estimated by the integer binary-search protocol as the largest passing load for which $\overline r_f\ge0.95$. This continuous-overlap criterion measures recovery or retention from a strong cue and does not require error-free correction of every flipped oscillator. All curves use the same initialization, observation time, and threshold protocol. The red curve is the second-resonance model in Eq.~\eqref{eq:Kuramoto2} with $\lambda_2=0.5$; the $J$--$K$ curves use $(J,K)=(3,10)$ (orange), $(1,10)$ (green), and $(0,10)$ (blue). Nominal fits over the displayed range give $\gamma_{\rm eff}=0.92$, $0.77$, $1.05$, and $2.00$, respectively, in the order just listed. The brackets are binary-search resolution intervals and the slopes are descriptive finite-size summaries, not asymptotic exponents.}
    \label{fig:ScalingPlot_zeroT}
\end{figure}

Let $r_\mu^{(a)}(t_f)$ be the terminal target overlap in trial $a$. The quantity used in Figs.~\ref{fig:ScalingPlot_finiteT} and \ref{fig:ScalingPlot_zeroT} is the mean terminal overlap
\begin{equation}
\overline r_f(P;N)
=
\frac{1}{N_{\rm trial}}
\sum_{a=1}^{N_{\rm trial}}r_\mu^{(a)}(t_f).
\label{eq:mean-terminal-overlap}
\end{equation}
This continuous statistic is not a retrieval success probability and no binary phase readout is applied. A complementary trial-level statistic is
\begin{equation}
p_{\rm succ}(P;N;r_*)
=
\frac{1}{N_{\rm trial}}
\sum_{a=1}^{N_{\rm trial}}
\mathbf 1\!\left\{r_\mu^{(a)}(t_f)\ge r_*\right\},
\label{eq:success-probability}
\end{equation}
which distinguishes a uniformly intermediate overlap from a mixture of high-overlap retrievals and low-overlap failures. The present figures were generated from Eq.~\eqref{eq:mean-terminal-overlap}; Eq.~\eqref{eq:success-probability} is stated to make the distinction explicit.

For a prescribed threshold $\overline r_c$, the empirical finite-time threshold $P_c(N)$ is located by an integer binary search in $P$. Starting from lower and upper loads chosen to bracket the transition, we evaluate the midpoint $P_{\mathrm{mid}}=\left\lfloor\frac{P_{\mathrm{low}}+P_{\mathrm{high}}}{2}\right\rfloor$  using $N_{\rm trial}=100$ fresh trials. If $\overline r_f(P_{\mathrm{mid}};N)\ge\overline r_c$, the midpoint becomes the new lower bound; otherwise it becomes the new upper bound. The search terminates when $P_{\mathrm{high}}=P_{\mathrm{low}}+1$, so that the final bracket consists of the largest passing value $P_{\mathrm{low}}$ and the smallest failing value $P_{\mathrm{high}}$. We report $P_c(N)=P_{\mathrm{low}}$.  This search assumes that the mean terminal overlap is approximately non-increasing with $P$ over the bracketing interval. The final binary-search bracket is an algorithmic load resolution, not a statistical confidence interval.

The finite-temperature protocol requests pre-jitter overlap $\epsilon_0=0.9$ and uses $\overline r_c=0.85$, so it measures preservation of a strong cue over the observation window rather than completion from a weaker cue. The zero-temperature protocol requests $\epsilon_0=0.95$ and uses $\overline r_c=0.95$. For a cue whose exact pre-jitter overlap is $0.95$, the expected overlap immediately after the jitter is $0.95\operatorname{sinc}(0.1)\simeq0.9484$. Thus the zero-temperature criterion tests recovery to or retention of a high-overlap state after a controlled off-binary perturbation; it does not require correction of every deliberately flipped oscillator and should not be interpreted as error-free pattern completion. The operational threshold depends on the cue construction, $\delta_{\rm init}$, $T_{\rm sim}$, and $\overline r_c$.

The dashed power laws summarize the displayed finite-size range. We do not attach formal statistical errors to their slopes: the plotted brackets record only the binary-search resolution. Larger systems and resampling over patterns, cues, and noise would be needed to estimate an asymptotic exponent.

The finite-temperature threshold simulations use $\Delta t=10^{-3}$ and $T_{\rm sim}=150$. Representative runs with $\Delta t=10^{-4}$ and $T_{\rm sim}=200$ showed no systematic change in the recorded observables or terminal overlaps. At nonzero temperature these remain terminal samples from a fluctuating state, not fixed points.

Figure~\ref{fig:ScalingPlot_finiteT} shows that the quartic-dominated models have larger finite-time strong-cue thresholds than the weakly quartic case over $20\lesssim N\lesssim300$, and several fitted slopes exceed one. These are finite-size trends rather than asymptotic predictions: for fixed $J>0$, the leading term in Eq.~\eqref{eq:sigma2} already reaches order unity at $P=\mathcal O(N)$. The $C=3$ tricritical point controls the single-pattern landscape, not the finite-load threshold slope.

Over the plotted range $20\lesssim N\lesssim150$, the purely quartic model has the largest operational threshold and a nominal slope near two. The ordering of the remaining curves is parameter-dependent, so the data do not support a blanket statement that adding a four-body channel improves every mixed model relative to the second-resonance baseline. The cavity variance estimate for $J=0$ identifies an $\mathcal O(N^3)$ signal--to--noise crossover scale obtained by formal extrapolation beyond the fixed-$\alpha$ regime of the closure, not an $N^3$ threshold law; its difference from the observed effective slope near two remains unresolved and emphasizes that the scalar closure does not determine the finite-size data. All zero-temperature curves use $\delta_{\rm init}=0.1$; Fig.~\ref{fig:ScalingPlot_zeroT} does not test sensitivity to this choice.

The criteria in Figs.~\ref{fig:ScalingPlot_finiteT} and \ref{fig:ScalingPlot_zeroT} fix a nonzero allowed error and use strong initial cues. They are therefore not directly comparable with the $\mathcal O(N/\log N)$ result of Ref.~\cite{nishikawa2004oscillatory}, which concerns perfect retrieval with an error tolerance that vanishes as $N\to\infty$. Figure~\ref{fig:ScalingPlot_zeroT} does provide a finite-$N$ comparison under a common operational protocol for the models shown, but a strict asymptotic comparison under a common error criterion remains open.

The single-pattern equilibrium phase diagram remains useful for selecting qualitative operating regimes. Below $C=3$ the onset is continuous, whereas above $C=3$ the equilibrium landscape can be bistable and requires a finite cue to cross the incoherent-side barrier. This low-load phase diagram describes accessibility and retention; it does not determine finite-load capacity.

\section{Analytical details of equilibrium bistability}\label{sec:equilibrium-details}
\subsection{Saddle--node line}\label{subsec:SN}
We now construct the saddle-node line for $C\equiv K/J>3$.
Define $X(x,r;C)=xr+\tfrac{Cxr^{3}}{6}$, where $x\equiv\beta J$. Fixed points satisfy
\begin{equation}
\label{eq:FP_compact}
r=\frac{I_{1}(X)}{I_{0}(X)}\equiv g(X),
\end{equation}
and a saddle--node (fold) occurs when
\begin{equation}
\label{eq:SN_condition}
\frac{\mathrm d}{\mathrm dr}\,g\bigl(X(x,r;C)\bigr)=1 .
\end{equation}
Solving Eqs.~\eqref{eq:FP_compact}--\eqref{eq:SN_condition} gives $(x_{\rm SN}(C),r_{\rm SN}(C))$. Figure~\ref{fig:oscmemory}(a) shows the stable and unstable branches; their intersection marks the fold.

\paragraph{Cusp at $C{=}3$.}
Expanding the Bessel ratio
$I_{1}(X)/I_{0}(X)=\tfrac12X-\tfrac1{16}X^{3}+\tfrac1{96}X^{5}-\tfrac{11}{6144}X^7+O(X^9)$ and using Eq.~\eqref{eq:FP_compact} yields, after retaining the terms that contribute through $O(r^{6})$ and dividing by $r(1-x/2)$, the polynomial
\begin{eqnarray}
0&\approx 1& - l_0 r^2 + m_0\biggl(\frac{x^2}{48}- \frac{C}{16}\biggr)r^4\nonumber \\
&+& m_0\biggl( \frac{5 C x^2}{288} - \frac{11 x^4}{3072}-\frac{C^2}{96} \biggr) r^6
\label{polyr2}
\end{eqnarray}
where $l_{0}=\frac{x}{x-2}\Bigl(\frac{x^{2}}{8}-\frac{C}{6}\Bigr)$ and $m_0=x^3/(x-2)$.

Writing $C=3+\delta$ ($\delta>0$), $x=2-\varepsilon$, and $y=r^2$,
the leading coefficients in Eq.~\eqref{polyr2} expand as
\[
l_0=\frac{\delta}{3\varepsilon}+1+O(\delta,\varepsilon),
\qquad
m_0=-\frac{8}{\varepsilon}+O(1),
\]
while
\[
\frac{x^2}{48}-\frac{C}{16}
=-\frac{5}{48}-\frac{\varepsilon}{12}-\frac{\delta}{16}+O(\varepsilon^2).
\]
Anticipating $\varepsilon=O(\delta^{2})$ (confirmed below), the quartic
coefficient is dominated by the product of the two leading terms,
$m_0\times\bigl(-\tfrac{5}{48}\bigr)=\tfrac{5}{6\varepsilon}$, so the leading
balance in Eq.~\eqref{polyr2} is
\[
0
=
1
-\frac{\delta}{3\varepsilon}\,y
+\frac{5}{6\varepsilon}\,y^2,
\]
with relative corrections of $O(\delta)$; the $r^6$ term of
Eq.~\eqref{polyr2} likewise enters only at $O(\delta)$. A non-degenerate
saddle-node requires this quadratic in $y$ to have a double root,
$\bigl(\tfrac{\delta}{3\varepsilon}\bigr)^{2}=4\cdot\tfrac{5}{6\varepsilon}$,
which gives
\[
\varepsilon=\frac{\delta^{2}}{30}+O(\delta^{3}),
\qquad
y=r^2=\frac{\delta}{5}+O(\delta^2).
\]
Keeping the next order in the same expansion gives

\begin{equation}
\label{eq:rc-small}
r_{\rm SN}(C)=\sqrt{\frac{\delta}{5}}\;+\;O(\delta^{3/2}), \quad
x_{\rm SN}(C)=2-\frac{\delta^{2}}{30}+O\bigl(\delta^{3}\bigr).
\end{equation}

The absence of a linear term in $x_{\rm SN}-2$ identifies $(C,x,r)=(3,2,0)$ as a codimension-two cusp: the fold is tangent to $x=2$, while $r_{\rm SN}$ grows as $(C-3)^{1/2}$.

\paragraph{Asymptotes for $C\to\infty$.}
When $C\gg1$ the cubic term in $X$ is dominant, so any finite
fixed-point solution must satisfy $xr=O(C^{-1})$.
Set
\begin{equation}
x=\frac{\kappa}{C},\quad r\rightarrow r_{\infty}\in(0,1),\quad
X\rightarrow X_{\infty}=\frac{\kappa}{6}\,r_{\infty}^{3},
\label{largeC1}
\end{equation}
where $\kappa=O(1)$  to be determined.
Insert Eq.~\eqref{largeC1} into the fixed-point condition Eq.~\eqref{eq:FP_compact}
and into the tangency condition Eq.~\eqref{eq:SN_condition};
the latter reads
$g'(X_{\infty})\,\bigl(x+\tfrac{Cx\,r_{\infty}^{2}}{2}\bigr)=1$.
Because $x=\kappa/C$ the $x$ term is negligible and one obtains
$g'(X_{\infty})\,\tfrac{C}{2}x r_{\infty}^{2}=1$,
i.e. $g'(X_{\infty})\,\kappa r_{\infty}^{2}/2=1$.

Using the Bessel identity
\[
g'(X)=1-\frac{g(X)}{X}-g(X)^2
\]
and the fixed-point relation $g(X_{\infty})=r_{\infty}$ gives

\begin{equation}
\left(1-\frac{r_\infty}{X_\infty}-r_\infty^2\right)
\frac{\kappa r_\infty^2}{2}=1.
\end{equation}

Since $X_\infty=\kappa r_\infty^3/6$, this becomes

\begin{equation}
\left(1-r_\infty^2-\frac{6}{\kappa r_\infty^2}\right)
\frac{\kappa r_\infty^2}{2}=1,
\end{equation}
hence,
\begin{equation}
\kappa=\frac{8}{r_{\infty}^{2}(1-r_{\infty}^{2})}.
\label{largeC2}
\end{equation}
Equations~\eqref{largeC1} and \eqref{largeC2} form the closed system
\begin{equation}
r_{\infty}=g\bigl(X_{\infty}\bigr),\quad
X_{\infty}=\tfrac{\kappa}{6}r_{\infty}^{3},\quad
\kappa=\tfrac{8}{r_{\infty}^{2}(1-r_{\infty}^{2})},
\end{equation}
which involves a single unknown $r_{\infty}\in(0,1)$.
Solving numerically yields
\begin{equation}
r_{\rm SN}(C)\xrightarrow[C\to\infty]{}r_{\infty}\simeq0.77733,\qquad
x_{\rm SN}(C)\sim\frac{33.454}{C}.\;
\label{eq:LargeC_result}
\end{equation}
Thus $x_{\rm SN}$ tends to zero as $C^{-1}$, while the fold overlap approaches $0.77733$. Strong quartic coupling shifts the fold but does not remove the finite-amplitude cue required to cross it; see Fig.~\ref{fig:oscmemory}(b).

\paragraph{Hyper-Catalan series for unstable branch} For an analytic approximation to the unstable branch at intermediate $C$, set $\zeta=l_0r^2$. Equation~\eqref{polyr2} then takes the cubic form used in the hyper-Catalan expansion of Ref.~\cite{wildberger2025hyper}; $\zeta$ is unrelated to the load $\alpha=P/N$:
\begin{equation}
    0=1-\zeta + t_2 \zeta^2 + t_3 \zeta^3,
\end{equation}
with
\begin{align}
t_{2}&=\frac{12x(x-2)(x^{2}-3C)}{(4C-3x^{2})^{2}},\label{eq:t2_def}\\[2pt]
t_{3}&=\frac{3(x-2)^{2}\bigl(96C^{2}-160Cx^{2}+33x^{4}\bigr)}{2\bigl(4C-3x^{2}\bigr)^{3}}.\label{eq:t3_def}
\end{align}
The cubic equation in $\zeta$ comes from the fixed-point expansion through $r^6$. Truncating its multivariate hyper-Catalan root at total degree three in $(t_2,t_3)$ gives
\begin{eqnarray}
\label{eq:rHC_stream_new}
r_u^{2}&=&
l_0^{-1}(
1+t_{2}+2t_{2}^{2}+5t_{2}^{3}
+t_{3}+5t_{2}t_{3}+21t_{2}^{2}t_{3}\nonumber \\
&+&3t_{3}^{2}+28t_{2}t_{3}^{2}+12t_{3}^{3}).
\end{eqnarray}
In Fig.~\ref{fig:oscmemory}(a), the green curve shows the truncated series in Eq.~\eqref{eq:rHC_stream_new}. It reproduces the numerical unstable branch visually over the displayed interval away from the fold. Because no uniform remainder estimate is available and the truncation cannot resolve branch coalescence, we use it only as a local analytic approximation and not to determine fold values, coexistence constants, or large-$C$ limits.

\subsection{Free-energy barriers and escape-time scaling}
\label{sec:kramers}
The microscopic Langevin dynamics in Eq.~\eqref{eq:SDE_definition} is $N$-dimensional, so neither $\Phi(r)$ nor $F_{\rm aux}(r)$ by itself defines a one-dimensional Markov stochastic differential equation for $r$. The exact constrained rate function $\Phi$ nevertheless determines the leading equilibrium large-deviation cost. Since Eq.~\eqref{eq:stationary-potential-equality} holds at every extremum, the required barrier differences may be evaluated from the simpler auxiliary potential.

Inside the bistable strip $x_{\mathrm{SN}}(C)<x<2$, let $r_s$ be the nonzero memory minimum and $r_u$ the intervening radial maximum. The two directional barriers are
\begin{align}
 B_{0\to m}(C,x)
 &:=\Phi(r_u)-\Phi(0)
 =F_{\rm aux}(r_u)-F_{\rm aux}(0),\\
 B_{m\to0}(C,x)
 &:=\Phi(r_u)-\Phi(r_s)
 =F_{\rm aux}(r_u)-F_{\rm aux}(r_s).
 \label{eq:directional-barriers}
\end{align}
They coincide only on the coexistence line $\Phi(r_s)=\Phi(0)$.

The auxiliary curvatures plotted in Fig.~\ref{fig:barriers} are
\begin{align}
 \ell_0^{\rm aux}&=[F_{\rm aux}''(0)]^{-1/2},\nonumber\\
 \ell_s^{\rm aux}&=[F_{\rm aux}''(r_s)]^{-1/2},\qquad
 \ell_u^{\rm aux}=[|F_{\rm aux}''(r_u)|]^{-1/2}.
\end{align}
They classify the stationary branches but are not physical fluctuation widths. At a stationary point $r_*$ with $X_*=\beta U'(r_*)$, the exact and auxiliary curvatures obey
\begin{equation}
\begin{aligned}
 \Phi''(r_*)&=\frac{1}{g'(X_*)}-\beta U''(r_*),
 \\
 F_{\rm aux}''(r_*)&=\beta U''(r_*)g'(X_*)\Phi''(r_*).
 \end{aligned}
 \label{eq:curvature-relation}
\end{equation}
For a strictly stable nonzero minimum, the leading radial standard deviation scales as $[N\Phi''(r_s)]^{-1/2}$.

For reversible mean-field dynamics, the expected leading escape-time scaling is
\begin{align}
 \log \tau_{m\to0}&=N B_{m\to0}+o(N),\\
 \log \tau_{0\to m}&=N B_{0\to m}+o(N).
 \label{eq:escape-large-deviation}
\end{align}
This is the part of the Eyring--Kramers picture that is fixed by the free-energy density \cite{kramers1940brownian,griffiths1966relaxation}. We do not write a scalar prefactor: deriving it would require the effective collective-coordinate mobility, the microscopic saddle and Hessian determinants, and treatment of the global phase zero mode. In particular, the common rotation period $2\pi/\omega_0$ cannot serve as an attempt time, because $\omega_0$ is removed in the co-rotating equilibrium model.

\paragraph{Near the thermodynamic tricritical point.}
Write $C=3+\delta$, $x=2-\varepsilon$, and $y=r^2$, with $\delta>0$ and $\varepsilon=O(\delta^2)$. The exact constrained free energy in Eq.~\eqref{eq:constrained-free-energy} gives the leading dimensionless potential
\begin{equation}
 \Phi(y)=\frac{\varepsilon}{2}y-\frac{\delta}{12}y^2+\frac{5}{36}y^3+o(\delta^3).
 \label{eq:tricritical-potential}
\end{equation}
The fold is obtained from a double stationary root,
\begin{equation}
 \varepsilon_{\mathrm{SN}}=\frac{\delta^2}{30},\qquad
 y_{\mathrm{SN}}=\frac{\delta}{5},
\end{equation}
consistent with Eq.~\eqref{eq:rc-small}. Equal-depth coexistence instead gives
\begin{equation}
 \varepsilon_{\mathrm{coex}}=\frac{\delta^2}{40},\qquad
 y_u=\frac{\delta}{10},\qquad
 y_s=\frac{3\delta}{10}.
\end{equation}
Consequently, the common barrier at coexistence is
\begin{equation}
 B_{0\to m}=B_{m\to0}
 =\Phi(y_u)-\Phi(0)
 =\frac{\delta^3}{1800}+o(\delta^3).
 \label{eq:tricritical-barrier}
\end{equation}
The cubic dependence on $C-3$ is therefore correct, but the fold and coexistence coefficients are distinct. At the fold $B_{m\to0}=0$, whereas at the incoherent spinodal $x\to2^-$ one has $B_{0\to m}\to0$.

\paragraph{Large-$C$ limit.}
For $C\gg1$ and $x=\kappa/C$, the small-root hyper-Catalan truncation gives $r_u\simeq\sqrt{12/\kappa}$ only when $r_u$ remains small, equivalently well away from the fold. It is not uniform as $\kappa\downarrow\kappa_\infty\simeq33.454$. At the fold the stable and unstable branches must instead merge at $r_\infty\simeq0.77733$, as found in Eq.~\eqref{eq:LargeC_result}. The incoherent-side barrier on the fold approaches
\begin{equation}
 B_{0\to m}^{\mathrm{fold}}
 \longrightarrow
 \frac{\kappa_\infty}{8}r_\infty^4
 -\ln I_0\!\left(\frac{\kappa_\infty}{6}r_\infty^3\right)
 \simeq0.2443.
 \label{eq:largeC-barrier}
\end{equation}
This number concerns activation out of the incoherent minimum; it is not the memory-retention barrier.

Figure~\ref{fig:barriers} plots $B_{0\to m}$, not $B_{m\to0}$. At fixed $C$, increasing $x$ from the fold toward $2$ decreases $B_{0\to m}$ to zero but increases $B_{m\to0}$ from zero. The two barriers can likewise respond oppositely to changes in $C$, so a single contour plot cannot be interpreted simultaneously as a cue-accessibility map and a memory-lifetime map. At $(C,x)=(10,1.7)$, evaluation of the stationary values of Eq.~\eqref{F}, equivalently of the exact constrained function in Eq.~\eqref{eq:constrained-free-energy}, gives
\begin{equation}
 r_u\simeq0.4264,\qquad r_s\simeq0.7866,
\end{equation}
\begin{equation}
 B_{0\to m}\simeq0.0131,\qquad B_{m\to0}\simeq0.0323.
\end{equation}
For $N=500$, the memory-side exponential factor is therefore $\exp(NB_{m\to0})\simeq\exp(16.1)$. An absolute lifetime cannot be inferred without the prefactor and the microscopic dynamical normalization. Deterministic relaxation in the heterogeneous-frequency model must instead be computed from the OA rate in Eq.~\eqref{eq:OA-rate} at an actual OA fixed point.

The directional barriers can therefore be used to compare equilibrium operating points. The OA relaxation rate remains a separate dynamical quantity.

\section*{Conclusions}

We studied a Hebbian oscillator memory with a two-body first-Fourier-harmonic interaction and a genuine four-body interaction. The four-body force can be evaluated in $\mathcal O(PN)$ operations per time step without constructing an $N^4$ tensor.

For identical oscillators with Langevin noise, the exact constrained rate function has a tricritical point at $\beta J=2$ and $K/J=3$. Near this point, $2-x_{\rm SN}\sim(C-3)^2/30$ and $2-x_{\rm coex}\sim(C-3)^2/40$, while the coexistence barrier is $(C-3)^3/1800$ to leading order. In the strong-quartic fold limit the incoherent-side barrier approaches $0.2443$. The two escape directions have different barriers, and the leading equilibrium prediction is $\log\tau\sim NB$. For the separate Lorentzian-frequency model, the Ott--Antonsen onset changes character at $K/J=6$; the identification $\Delta=1/\beta$ applies only to the linear threshold.

At finite load we used a cavity signal--to--noise closure with stated weak-dependence and projection assumptions. It gives a crossover scale $P=\mathcal O(N)$ for fixed $J>0$; formal extrapolation of its quartic term gives $P=\mathcal O(N^3)$ for $J=0$, outside the fixed-$\alpha$ regime of the closure and hence not a controlled capacity prediction. Neither scale determines the measured threshold slopes. The simulations use strong cues and fixed continuous-overlap criteria. Across the simulated parameter sets, the largest thresholds occur in quartic-dominated cases, and the pure-quartic zero-temperature data have the largest nominal slope. These are finite-size recovery and retention results, not asymptotic capacity laws or error-free completion capacities.

The cited photonic, polaritonic, superconducting, electronic, and spintronic systems provide ingredients rather than complete realizations of the model. Implementing Eq.~\eqref{eq:high-order-kuramoto} would also require programmable signed Hebbian weights. The results connect higher-order Kuramoto dynamics with dense associative memory, while a controlled finite-load theory and larger-scale statistical tests are still needed to establish asymptotic capacity.

\section*{Acknowledgments}
J.N. acknowledges the support from the EPSRC through PhD studentship EP/W524633/1. N.G.B. acknowledges support from the HORIZON EIC Pathfinder Challenges project HEISINGBERG (grant 101114978), the Weizmann--UK Make Connection grant (grant 142568), and the EPSRC UK Multidisciplinary Centre for Neuromorphic Computing (grant UKRI982).

\section*{Data Availability}
The numerical data and code supporting the findings of this study are available from the corresponding author upon reasonable request.

\bibliography{meanfield,literatureNew}

\end{document}